\documentstyle[aas2pp4,twoside]{article}
\input epsf

\newcommand{\oi}{O~{\sc i}}

\newcommand{\Jtco}{J=1$\rightarrow$0~$^{13}$CO}

\newcommand{\arcs}{{$^{\prime \prime}$}}
\newcommand{\arcm}{{$^{\prime}$}}
\newcommand{\kms}{km~s{$^{-1}$}}

\newcommand{\msol}{M{$_{\odot}$}}
\newcommand{\lsol}{L{$_{\odot}$}}

\begin{document}
\title{Supersonic Random Flows in the Perseus Molecular Cloud}

\author{Paolo Padoan}
\affil{Theoretical Astrophysics Center, 
       Juliane Maries Vej 30, DK-2100 Copenhagen, 
       Denmark}

\author{John Bally \& Youssef Billawala}
\affil{Department of Astrophysics, Planetary, and Atmospheric Sciences,\\
       Center for Astrophysics and Space Astronomy,\\
       Campus Box 389, University of Colorado, Boulder CO 80309} 

\author{Mika Juvela}
\affil{Helsinki University Observatory, 
       T\"ahtitorninm\"aki, P.O.Box 14,
       SF-00014 University of Helsinki, Finland}

\author{\AA ke Nordlund}
\affil{Astronomical Observatory and Theoretical Astrophysics Center, \\
       Juliane Maries Vej 30, DK-2100 Copenhagen, Denmark}

\authoremail{padoan@tac.dk}

\begin{abstract}

We compare the statistical properties of J=1$\rightarrow$0 $^{13}$CO
spectra observed in the Perseus Molecular Cloud with synthetic
J=1$\rightarrow$0 $^{13}$CO spectra from a 5~pc model cloud. The synthetic
spectra are computed solving the non--LTE radiative transfer problem for a
model cloud obtained as solutions of the 3--D magneto--hydrodynamic (MHD)
equations in both the highly supersonic and super--Alfv\'{e}nic regimes of
random flows. 

We present several statistical results that demonstrate remarkable
similarity between real data and the synthetic cloud. The
three--dimensional structure and dynamics of molecular clouds like Perseus
are appropriately described by random supersonic and super--Alfv\'{e}nic
MHD flows. Although the description of gravity and stellar radiation are
essential to understand the formation of single protostars and the effects
of star formation in the cloud dynamics, the overall description of the
cloud and of the initial conditions for star formation can apparently be
described on intermediate scales without accounting for gravity and
stellar radiation.

\end{abstract}


\section{Introduction}

The structure and dynamics of molecular clouds (MCs) has been the subject
of intensive investigation during the past two decades. Since MCs are the
sites of star formation, their properties provide insights into the
initial conditions needed to form stars. On the other hand, energy
injected into the clouds by young stars is also believed to play a crucial
role in the cloud dynamics and evolution (e.g. Whitworth 1979; Bally \&
Lada 1983; Miesch \& Bally 1994). The observed state of a cloud represents
the balance between energy injection and dissipation. 

Observations show that cloud internal motions are highly supersonic,
incoherent, and random (e.g. Bally et al. 1987, 1989, 1990).
Furthermore, the cloud structure is complex and by many measures is
suggestive of the presence of random flows. Indeed, many authors have
attempted to describe clouds using various models of turbulence (e.g.
Ferrini, Marchesoni \& Vulpiani 1983; Henriksen \& Turner 1984; Scalo
1987; Fleck 1988; Fleck 1996) inspired by the scaling laws observed in the
interstellar medium (Larson 1981). More recently, there have been several
attempts to describe cloud morphology and velocity fields using fractals
(e.g. Scalo 1990; Falgarone \& Phillips 1991; Falgarone, Phillips \&
Walker 1991; Falgarone 1992; Larson 1995; Elmegreen 1997). 

Most theoreticians have argued that clouds are supported by magnetic
fields (Mestel 1965; Strittmatter 1966; Parker 1973; Mouschovias 1976a,
b; McKee \& Zweibel 1995) and interpreted the observed line--widths in terms
of magneto--hydrodynamic (MHD) waves (Arons \& Max 1975; Zweibel \&
Josafatsson 1983; Elmegreen 1985, Falgarone \& Puget 1986). However,
comparison of the extinction statistics with recent models of supersonic
and super--Alfv\'{e}nic random flows suggests that such models may provide
a superior description of MCs (Padoan, Jones \& Nordlund 1997). OH
Zeeman measurements (Crutcher et al. 1993; Crutcher et al. 1996) have also
been shown to be consistent with the predictions of a super--Alfv\'{e}nic
random flow (Padoan \& Nordlund 1997). 

In this paper, we compare some observed properties of the Perseus
molecular cloud with the predictions of supersonic and super--Alfv\'{e}nic
random flows. We start with a self--consistent MHD simulation of a random
flow in a 3--D $128^3$ grid (Padoan \& Nordlund 1997). We solve the
radiative transfer problem using a non--LTE Monte Carlo approach (Juvela
1997) to compute $^{13}$CO synthetic spectra on a 90 by 90 cell grid with
60 velocity channels. Remarkable agreement of these models with the data
is illustrated by several results. These results demonstrate that such
models provide an excellent description for the kinematics and structure
of molecular clouds despite the absence of gravity, external radiation
fields, or stellar outflows in the models. 

\section{The Perseus Molecular Cloud}

Observations of the Perseus molecular cloud complex were made from 1985 to
1991 with AT\&T Bell Laboratories 7~m offset Cassegrain antenna in
Holmdel, New Jersey. Approximately a 6$^\circ$ by 4$^\circ$ region was
surveyed in the 110 GHz line of \Jtco . The observations were made on a
1\arcm\ grid with a 100\arcs\ beam yielding 33,000 spectra, each with
128~$\times$~100~kHz channels (0.273~\kms ). See Billawala et al. (1997)
for further details of observations and discussion of the cloud structure. 

The Perseus molecular cloud is roughly a 6$^\circ$ by 2$^\circ$ region
located at a distance estimated to range from 200 to 350~pc (Herbig \&
Jones 1983; C\^ernis 1990) and has a projected size of 30 by 10~pc (see
figure 1). The cloud is part of the larger Taurus--Auriga--Perseus complex
and lies below the galactic plane ({\it b} $\sim\ -20^\circ $) with an
estimated mass of 10$^4$~\msol\ (Billawala et al. 1997). 

The complex has produced some high mass stars at its eastern end which lie
near the centroid of Perseus OB2 association, a loose grouping of 5 to 10
O and B stars, that is about 7 million years old (Blaauw 1991). Perseus
OB2 has blown a 20$^\circ$ diameter (100~pc) supershell into the
surrounding interstellar medium that can be seen in 21~cm H~${\sc i}$ data
(Heiles 1984). The shell is partially superimposed on the Perseus
molecular cloud and may be interacting with it. Perseus contains two young
stellar clusters and a background of relatively inactive molecular gas
that has formed few stars. 

Near the eastern end is the young ($<$ 7~Myr old) cluster IC348, which has
several hundred members and may still be forming (Strom, Grasdalen \&
Strom 1974; Lada, Strom \& Myers 1993; Lada \& Lada 1995). Omicron Persei,
a B1 III star, located near the center, is believed to be a few parsecs 
in front of the cloud
(Bachiller et al. 1987). Lada \& Lada (1995) estimate a cluster population
of $\approx $ 380 stars with a highly centralized distribution falling off
inversely with distance from the inner 0.1~pc diameter core which contains
about $\sim$~200~\msol\ of stars and a density of 220~\msol~pc$^{-3}$.
Star formation has been going on for at least 5 to 7~Myr at a roughly
constant rate and with an efficiency of $\sim~50\%$. 

A second and even younger cluster is embedded in the Perseus cloud several
degrees to the west near the reflection nebula NGC1333. This region
contains the most active site of ongoing star formation in Perseus.
NGC1333 contains an embedded $<$~1~Myr old infrared cluster, a large
number of molecular outflows (Knee \& Sandell 1997), Herbig--Haro objects
(Bally, Devine \& Reipurth 1996), and shock excited near--infrared H$_2$
emission regions (Aspin, Sandell \& Russell 1994; Hodapp \& Ladd 1995; 
Sandell et al. 1994). Lada, Alves \& Lada (1996) find 143 young stars 
in a 432 square
arc--minute region with most being members of two prominent sub--clusters. 
More than half of these stars display infrared excess emission, a
signature of young stellar objects. Lada et al. estimate the age of the
cluster to be less than $10^6$ years with a star formation rate of
$4~\times~10^{-5}$~\msol~yr$^{-1}$. 

Away from IC348 and NGC1333, the Perseus cloud contains about a dozen
dense cloud cores with low levels of star formation activity including the
dark cloud B5 (Langer 1989; Bally, Devine, \&  Alten  1996) at its eastern
end, the cold core B1 for which a magnetic field has been measured by
means of OH Zeeman (Goodman et al. 1989; Crutcher et al. 1993), and the
dark clouds L1448 and L1455 on the west end (Bally et al. 1997). 

The cloud exhibits a wealth of sub--structures such as cores, shells, and
filaments and dynamical structures including stellar outflows and jets,
and a large scale velocity gradient. While emission at the western end of
the complex lies mostly near v$_{lsr}$~=~4~\kms\ (L1448), the eastern end
of the complex has a velocity of v$_{lsr}$~=~10.5~\kms (B5). This may be
an indication that we are seeing several smaller clouds partially
superimposed along the line of sight (C\^ernis 1990). 

\section{Theoretical Models}

Padoan et al. (1997) computed synthetic molecular clouds containing grids
of 90~$\times$~90 spectra of different molecular transitions for a total of
more than a half a million spectra. The spectra were obtained with a
non--LTE Monte Carlo code (Juvela 1997) starting from density and velocity
fields that provide realistic descriptions of the observed physical
conditions in MCs. The density and velocity fields are obtained as
solutions of the magneto--hydrodynamic (MHD) equations in a 128$^3$ grid
and in both super--Alfv\'{e}nic and highly supersonic regimes of random
flows (Padoan \& Nordlund 1997). The resulting density fields span a
continuous range of values from 0.1 to 10$^5$ cm$^{-3}$ which produce
column densities ranging over three orders of magnitude or more. These
cloud models have already been shown to reproduce observed statistical
properties of MCs (Padoan, Jones \& Nordlund 1997; Padoan \& Nordlund
1997). 

In this work we use a 90~$\times$~90 grid of spectra of \Jtco\ from a 5~pc
diameter artificial cloud. For a detailed description of the construction
of the synthetic spectra we refer the reader to Padoan et al. (1997). 

The MHD numerical experiment which produces the synthetic cloud does not
account for gravity, stellar radiation, or stellar outflows. However, the
numerical flow is consistent with the general observed properties of dark
clouds such as the existence of random and supersonic motions. The mean
density and rms velocity in the model are related to the physical size of
the computational grid via Larson's interstellar medium scaling laws
(Larson 1981). A 5~pc model has been selected because the Perseus complex
looks like a string of smaller clouds with an estimated size of a few
parsecs. The mean density of the 5~pc model is 400 H$_2$ molecules
cm$^{-3}$. This corresponds to a total mass of $3.2 \times 10^3$~\msol\
(using a mean molecular weight of $\mu$~=2.6 to correct for He). 

\section{Statistics of the Observed and Synthetic Spectra }

Figure 1 shows the integrated \Jtco\ map of the Perseus complex. The two
brightest emission regions lie near the young stellar clusters IC348
(left) and NGC1333 (right). Complex and filamentary structure is evident
in this image, in the individual channel maps, and in images of peak
antenna temperature (Billawala et al. 1997). Figure~2 left and center
panels show close--up views of the L1448 and B5 dark clouds respectively. 
Figure~2 on the right shows one particular synthetic map used in our
analysis. It is obtained from the 5~pc cloud model and shows the
integrated antenna temperature of the \Jtco\ synthetic spectra as the
three observed maps in figures 1 and 2. As the real data, the synthetic
image shows filamentary structure, cores, and partial shells. 

Fig.~3 shows the mean spectrum obtained by averaging all spectra lying
within the boundaries of the maps of each of three regions; B5, L1448, and
the entire Perseus complex. We over--plot the synthetic mean spectrum
computed from the 5~pc synthetic map (dashed lines). The spectra are
centered about their own mean. The high velocity blue--shifted tail in the
L1448 spectrum may be produced by an unrelated cloud superimposed along
the line--of--sight. The B5 cloud has been estimated to have a mass of
slightly less than $10^3$~\msol\ (Langer et al. 1989) while the mass of
L1448 has been estimated to be about 500~\msol\ by Bally et al. (1997). 
These results are consistent with the narrower width of the mean spectra
as compared to the width of the mean spectrum of the more massive
(3200~\msol ) theoretical model. The mean spectrum of the entire Perseus
cloud is wider than that of the theoretical model for different reasons. 
First, the total mass of the Perseus complex is about $10^4$~\msol\
(Billawala et al. 1997). Second, there is a large scale systematic
velocity gradient from the eastern to the western ends of the complex
which are not taken into account in our models. The projected length of
the cloud is 30~pc, far larger than the size of our model. Finally, there
is evidence that this gradient is produced by the superposition of two
clouds that may be as far as 100~pc distant from each other (C\^ernis
1990; Herbig \& Jones 1983). 

The probability distributions of the velocity integrated
antenna temperature for L1448, B5, and the entire Perseus cloud are plotted
in Fig.~4. The theoretical result derived from the 5~pc diameter model cloud is
over--plotted in dashed lines. Though there are individual differences
between the clouds and the model, the overall shapes of the distributions
are qualitatively similar in that they are dominated by intermittent
tails. Differences between L1448 and B5 may be in part due to
line--of--sight confusion and small number statistics. For the entire
Perseus cloud, the main difference may be attributable to the fact that it
is a much larger cloud than our model. The distributions of larger models
are in fact expected to be more intermittent (Padoan \& Nordlund 1997).
Since the integrated antenna temperature is proportional to the column
density, this result shows that the theoretical density distribution is
consistent with the observed one. 

Figure~5, 6, and 7 show 30 by 30 grids (30\arcm\ by 30\arcm\ for the real
data) of spectra for L1448, B5, and the model respectively. In all three
figures, we have selected the highest intensity regions of these three
maps. Note the qualitative similarity between the real data and the model
spectra. 

Figure~8 through 11 show histograms of the first four statistical moments
of the individual spectra for all of Perseus, L1448, B5, and the model
respectively. The plots in the upper left show the distributions of
velocity centroids centered around their means. The upper right plots
show the distributions of the line velocity dispersions (the line widths).
The lower left plots show the skewness of these profiles which measures
the degree of asymmetry present in the profiles. The observed histograms,
like the theoretical ones, present a distribution of skewness values
around 0, that is the gaussian value. They have a bit more extended
tails than the theoretical histograms, but unfortunately the tails are 
rather sensitive to the noise in the observed spectra, and there is no general
way to deal with the noise in order to make the tails statistically significant. 
The lower right plots show the kurtosis which measures the intermittency 
(high velocity tails) in the
spectra. The distribution of kurtosis for the theoretical cloud are
similar to the observed distributions. However, the observed peaks are
centered around 2.3 rather than 3.0 for the theoretical model. The
vertical dashed lines correspond to the values of the same statistical
moments in the mean spectra shown in figure~3. 

Scatter plots of the equivalent width versus velocity
integrated antenna temperature are shown in Fig.~12. 
The equivalent width is here defined as the velocity integrated 
antenna temperature divided by the peak antenna temperature in 
each spectrum. Note that the theoretical and real data agree 
in magnitude, in dispersion, and have a similar overall shapes.
The observed values of the line--width at the lowest integrated 
temperatures are a bit smaller than the theoretical ones, but that 
region of the plots should not be compared with the theoretical
results, because the contribution from the noise in the observed 
spectra is there too strong.  
B5 apparently has a smaller equivalent width for every value of
integrated temperature probably because it is a smaller cloud and 
is less dynamic than L1448. Over a half--dozen active outflows have 
been discovered in L1448 (Bally et al. 1997) while only two are
known in B5 (Bally, Devine \& Alten 1996).

\section{Discussion and Results}

The model we constructed was built upon very general considerations and
not tailored to the specific characteristics of the Perseus molecular
cloud or individual cores. Only the general interstellar medium scaling
laws (Larson 1981) have been used to fix the physical size of the cloud
model. The model has been inspired by the observation that random
supersonic motions are ubiquitous in molecular clouds. 

The dynamic range of length scales covered by the MHD calculations ranges
from 0.04~pc to 5~pc (128$^3$ grid--points). The radiative transfer 
calculations and the resulting spectra cover a linear scale range from 
0.05~pc to 5~pc (90$^3$ grid--points). However, numerical dissipation 
effectively degrades this
resolution by a factor of two to an inner scale of about 0.1~pc. Thus the
inner scale of the theoretical models are comparable to the linear
resolution of the 100\arcs\ beam used in our observations. The Bell Labs
7~meter telescope beam has a linear resolution of about 0.14 pc (assuming
a distance of 300~pc for the Perseus cloud) and the maximum extent of the
Perseus cloud that has been mapped is about 30~pc. 

Giant molecular clouds (GMCs; M$~>~10^5$~\msol\ and L $\approx$ 20 to
50~pc) are gravitationally bound. This must be the case since the product
of the internal velocity dispersion squared times the density is one to
two orders of magnitude greater than the pressure of the surrounding
interstellar medium. Gravity must provide the restoring force on the
scale of a GMC, otherwise GMCs would be dispersed before they could form
any stars. Individual stars form from the gravitational contraction of
small cores with sizes of order 0.1~pc, while clusters of stars such as
IC348 and NGC1333 form from cores about 1~pc in diameter. Thus gravity
is certainly important on the very large scale of GMCs (L $>$ 20~pc) and
on the very small scales of dense star forming cores (L $\le$ 1~pc). Our
model and observations have linear dimensions intermediate between these
two scales. The results of this work show that on intermediate scales the
observed properties of molecular clouds and likely the initial conditions
for star formation can be appropriately described without explicitly
accounting for gravity. Self--gravitating cores may occasionally condense
from the random flow where a local over-density is produced statistically.
However, at any one time, only a fraction of the total mass is involved in
such condensations. If indeed self--gravitating cores condense from the
intermittency of the random flow, then it might be possible to predict the
mass spectrum of the resulting condensations which produce stars. This may
be an important ingredient in constructing the initial stellar mass
function (Padoan, Nordlund \& Jones 1997). 

In this picture, the overall density structure inside a GMC is {\it not}
the result of gravitational fragmentation, but rather of the presence of
supersonic random motions together with the short cooling time of the
molecular gas. We predict that there are several shocks in the gas along
any line--of--sight with velocities comparable to the overall cloud
line--width. These shocks have velocities of order 1 to 10 \kms\ which are
not observable in the forbidden lines in the visible portion of the
spectrum. However, there are several far--infrared transitions such as the
157~$\mu$m fine--structure line of C$^+$, the 63~$\mu$m line of \oi \ , and
the 28, 17, and 12 $\mu$m v~=~0--0 lines of H$_2$ which are readily
excited by such shocks. The COBE and ISO satellites have shown that these
transitions together carry about 0.1\% of the total luminosity of a
typical galaxy (Wright et al. 1991; cf. Lord et al. 1996). 
We predict that
roughly 10\% of this emission is produced by the shocks discussed above
because these transitions are the primary coolants in the post--shock gas. 

The dissipation of the kinetic energy of internal motions, $E_{GMC}$, by
shocks in a GMC produces a luminosity of order $L_{GMC} ~\approx ~ E_{GMC}
/ \tau _{GMC} ~\approx $ $3 \times 10^3$~(\lsol\ )~$E_{51} \tau _{4 \times
10^6}$ , where the dynamical time is taken to be a cloud crossing time,
$\tau _{GMC} ~=~R_{GMC}/ \sigma _{GMC}$ and $\sigma _{GMC}$ is the cloud
velocity dispersion. Since the above infrared transitions are the main
coolants in the post--shock layers of these low velocity shocks, we expect
that future observations of clouds like Perseus will show extended bright
emission in these infrared transitions. 

Observations of young stellar populations show that stars form from GMCs
over a period of order 10 to 20 Myrs (Blaauw 1991) which implies that
clouds must survive for at least this long. Since this time--scale
exceeds the dissipation time, the internal motions must be regenerated. 
There are several possible sources for such energy generation. When
massive stars are present, their radiation, winds, and supernova
explosions inject large amounts of energy into the surrounding gas.
However this amount of energy is far in excess of what is required to
balance the dissipation of kinetic energy in shocks. Massive stars are
likely to be responsible for cloud disruption. 

The second candidate energy source is outflows from low mass stars which
form more uniformly throughout the cloud (Strom et al. 1989; Strom, 
Margulis \& Strom 1989). The Perseus
molecular cloud contains only intermediate to low mass stars, which during
their first $10^5$ years of life produced jets and outflows. All the cores
studied here are known to have multiple outflows. Over a dozen outflows in
NGC1333, 8 groups of Herbig--Haro objects in L1448, and 2 outflows in B5
have been discovered. Outflows can provide the mechanism to stir the
clouds and balance the dissipation of the random supersonic motions. 

The plots of equivalent width versus integrated antenna temperature in
figure 12 support the scenario described above. While the plot for L1448
is in excellent agreement with the model, the plot for B5 shows a
systematically lower equivalent width. This is consistent with the lower
number of outflows discovered in B5 relative to L1448. 

However, the main source of energy for the small and intermediate scale
differential velocity field may simply be cascading of energy from larger 
scales.  The dynamical times of intertial scale motions are short compared
to those of the larger scale motions. In the inertial range,
there is a quasi-stead flow of energy from larger scales towards the
energy dissipation scales.  The numerical simulations represents 
only piece of the inertial range, and hence, on these scales, energy
input at the largest wavenumbers is to be expected.

On the largest, energy carrying scales, we have already remarked that
the motions must be constrained by gravity.  Gravity field 
fluctuations large enough to constrain these clouds must come from  
self-gravity, and from interactions with neighboring clouds and 
embedded stars.  The time variations and tidal effects due to relative
motions of the self-gravitating masses is perhaps the main source
of energy input at the large scale end of the inertial cascade.

Padoan, Jones \& Nordlund (1997) and Padoan \& Nordlund (1997) have shown
that the statistics of infrared stellar extinction (Lada et al. 1994) and
the OH Zeeman measurements (Crutcher et al. 1993; Crutcher et al. 1996)
can be explained by our model of supersonic random flows. In the present
work, we have shown that the same model is consistent with molecular
clouds observed in the \Jtco\ line. These results together provide
compelling evidence that {\it supersonic and super--Alfv\'{e}nic random
flows correctly describe the structure and dynamics of molecular clouds.}

\acknowledgements

This work has been partially supported by the Danish National Research
Foundation through its establishment of the Theoretical Astrophysics
Center. Computing resources were provided by the Danish National Science
Research Council, and by the French `Centre National de Calcul
Parall\`{e}le en Science de la Terre'. PP is grateful to the Center for
Astrophysics and Space Astronomy (CASA) in Boulder (Colorado) for the warm
hospitality offered during the period in which this paper has been
written. JB and YB acknowledge support from NASA grant NAGW--4590 (Origins)
and NASA grant NAGW--3192 (LTSA). The work of MJ was supported by the
Academy of Finland Grant No. 1011055.

\clearpage

{\bf Figure and Table captions:} \\

{\bf Figure 1:} Integrated antenna temperature of \Jtco\ for velocities
0 to 15~\kms\  \\

{\bf Figure 2:} Integrated antenna temperature of the L1448 (left panel), 
B~5 (center panel), and the 5~pc model (right panel). As the real data,
the synthetic image shows filamentary structure, cores, and partial shells. \\

{\bf Figure 3:} Mean spectra of L1448 (upper panel), B5 (middle panel), and
the entire Perseus molecular cloud (lower panel). The dashed curves 
shows the mean spectrum of the model. Each spectrum is centered around its
own mean. \\

{\bf Figure 4:} Probability distribution of velocity integrated antenna
temperature for the three clouds as in figure 3. The dashed curve
shows the theoretical distribution.  \\

{\bf Figure 5:} A 30\arcm\ $\times$ 30\arcm\ map of individual \Jtco spectra 
around the brightest region of L1448. The velocity interval for each
spectrum shown is 6.5~\kms\ (the actual interval used for the statistical
analysis was 15\kms ) and the antenna temperature ranges from 0 to 6 K.  \\

{\bf Figure 6:} As in figure 5 but for the B5 cloud. \\

{\bf Figure 7:} As in figure 5 but for the 5~pc model cloud. \\

{\bf Figure 8:} Histograms of the first four statistical moments of the
spectra of the entire observed region of Perseus. Upper left: 
distribution of centroid velocities. Upper right: distribution of 
velocity dispersions (line widths). Lower left: distribution of 
skewness. Lower right: distribution of kurtosis. Vertical dashed
lines show the values of the same statistical moments for the mean 
spectrum.  \\

{\bf Figure 9:} As in figure 8 but for L1448. \\

{\bf Figure 10:} As in figure 8 but for B5. \\

{\bf Figure 11:} As in figure 8 but for the 5~pc model cloud. \\

{\bf Figure 12:} Scatter plots of equivalent width versus velocity integrated
antenna temperature for the three observed regions (three upper panels)
and for the 5~pc model cloud (lower panel). The diamond symbols show
the mean value of the equivalent width in each interval of integrated
antenna temperature and the "error bars" show the one $\sigma $ 
distribution around the mean. Intervals of integrated antenna
temperature were chosen to contain the same number of points. \\

\onecolumn

\newpage
\begin{figure}
\centering
\leavevmode
\epsfxsize=1.0
\columnwidth
\epsfbox{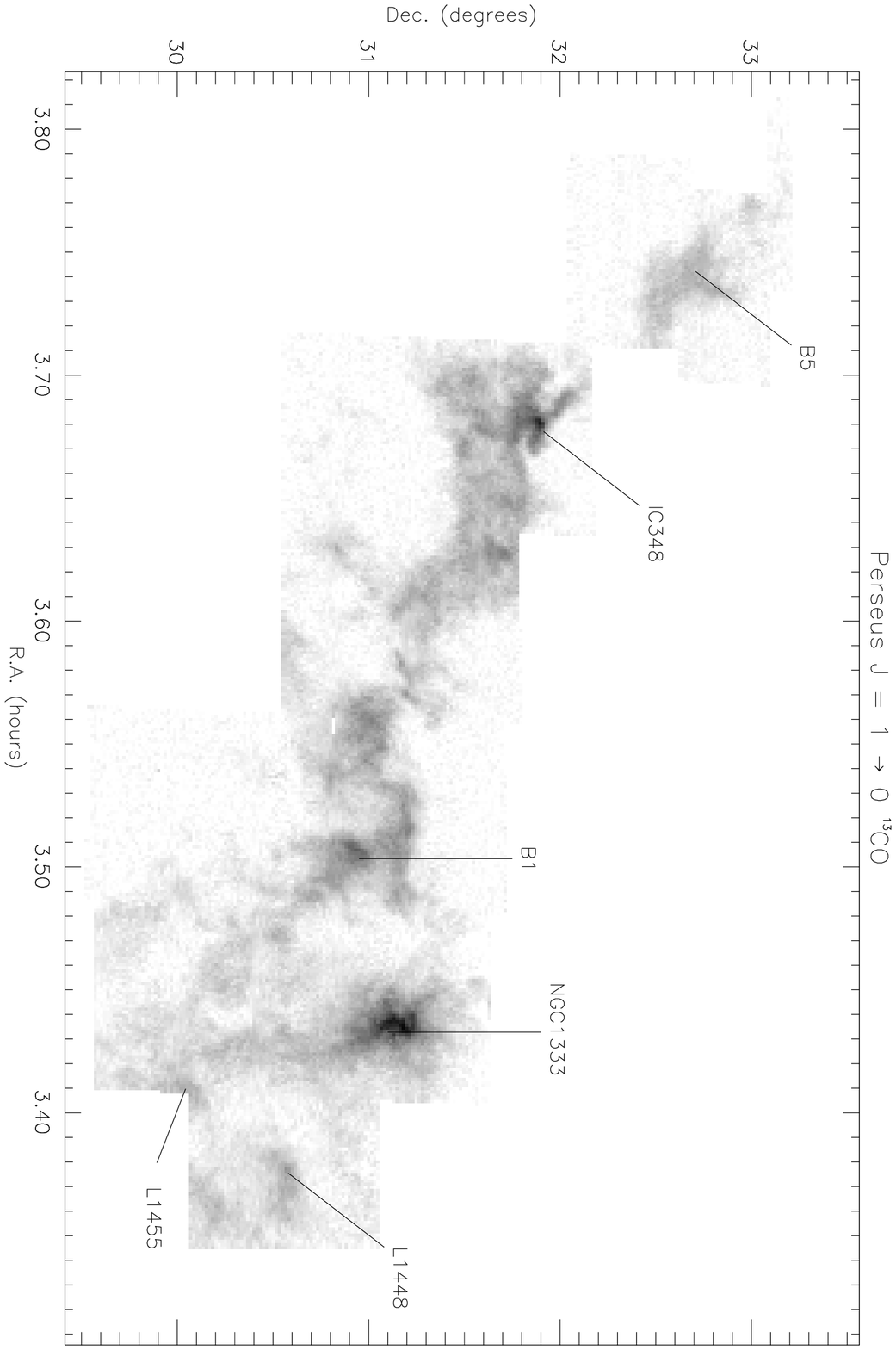}
\caption[]{}
\end{figure}

\newpage
\begin{figure}
\centering
\leavevmode
\epsfxsize=0.4
\columnwidth
\epsfbox{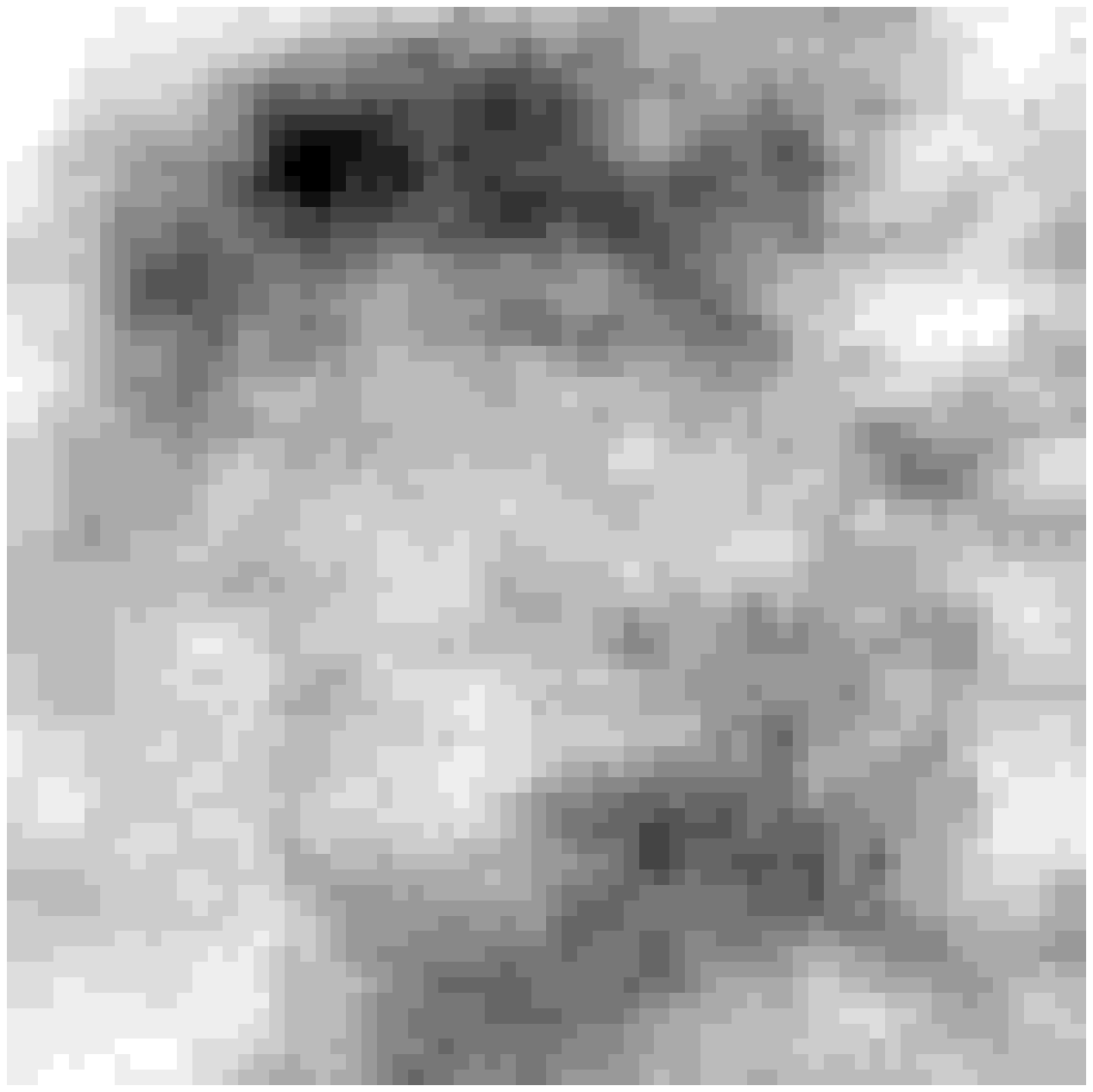}
\epsfxsize=0.4
\columnwidth
\epsfbox{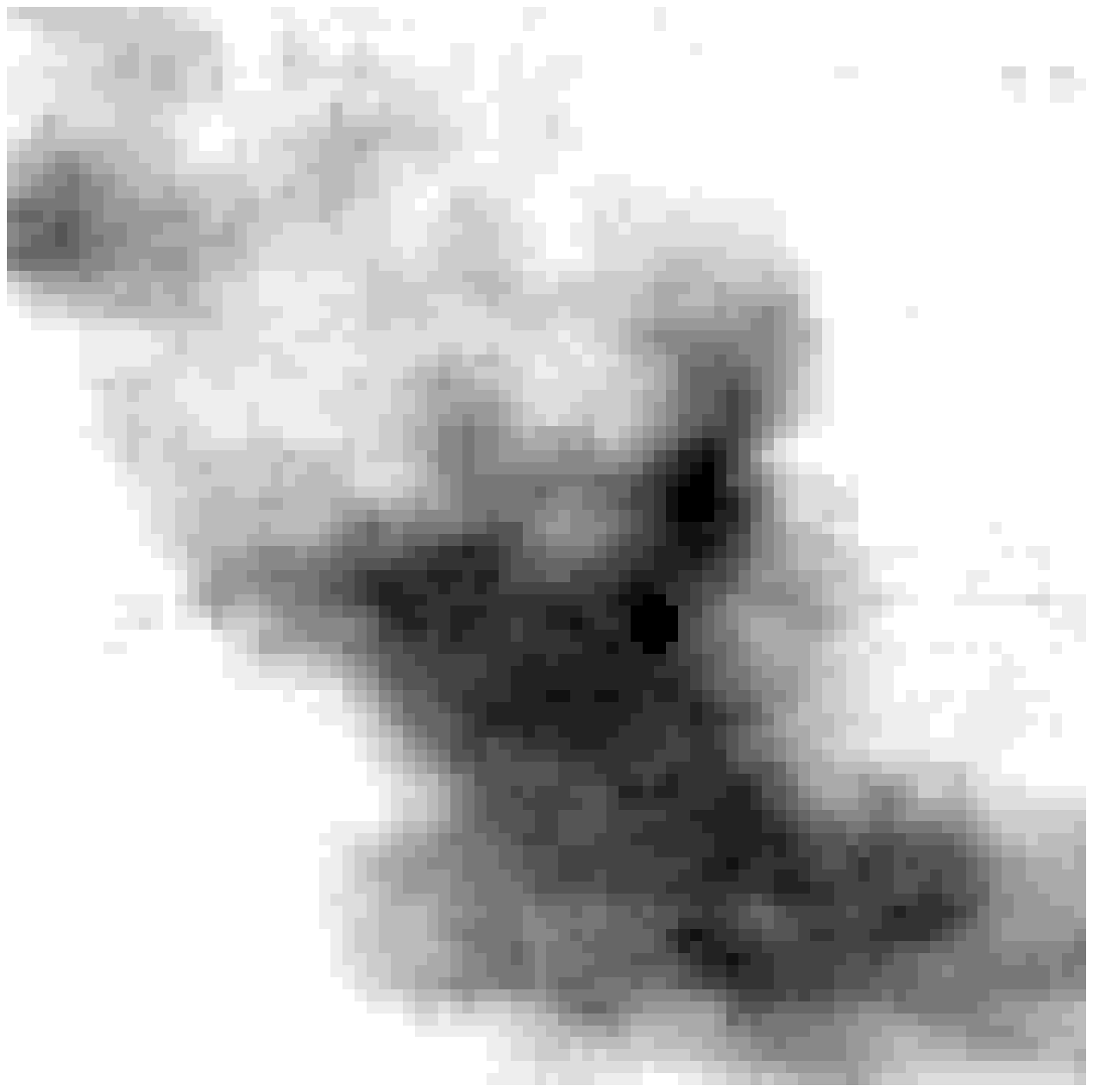}
\epsfxsize=0.4
\columnwidth
\epsfbox{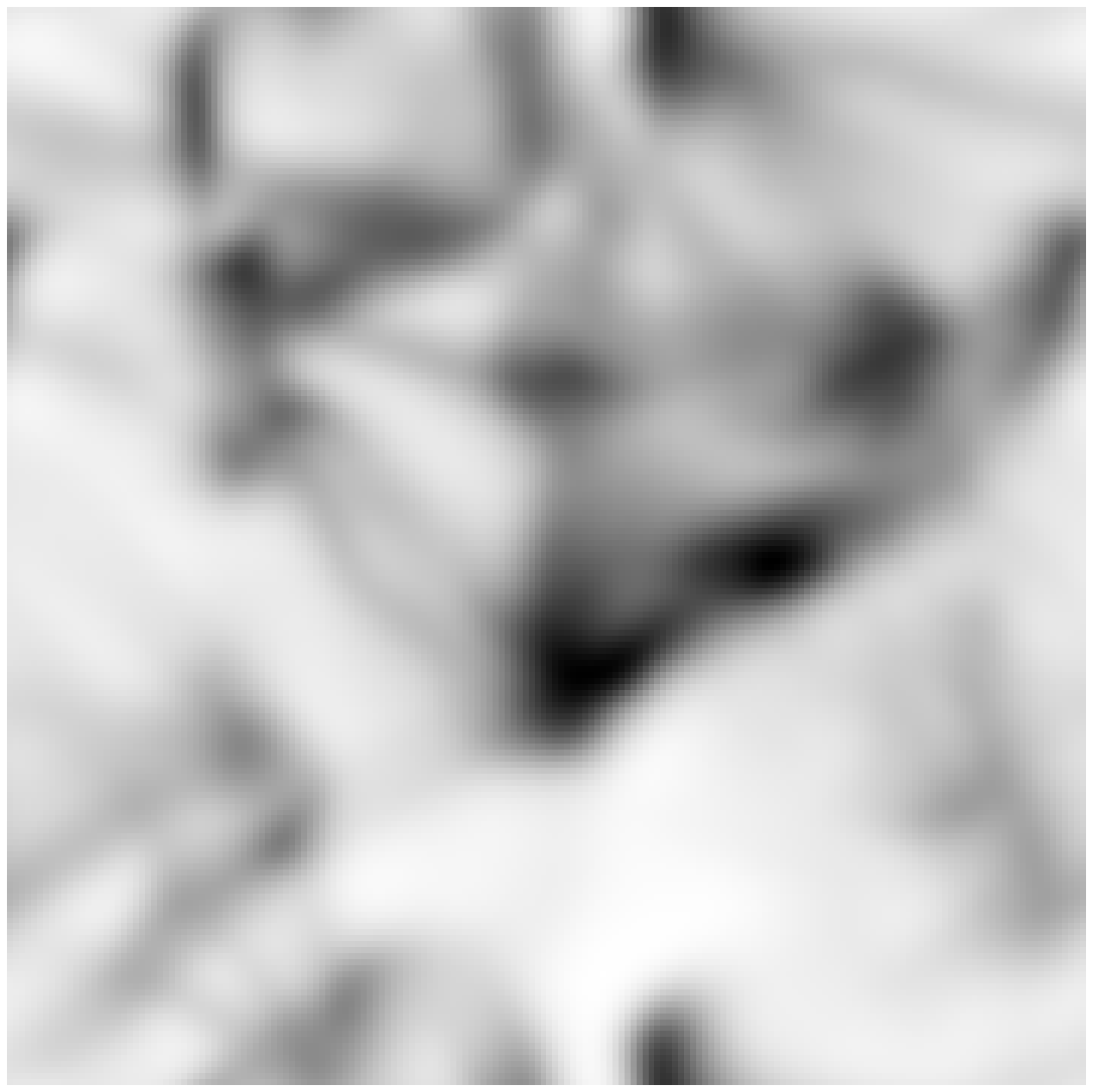}
\caption[]{}
\end{figure}

\newpage
\begin{figure}
\centering
\epsfxsize=0.5
\columnwidth
\epsfbox{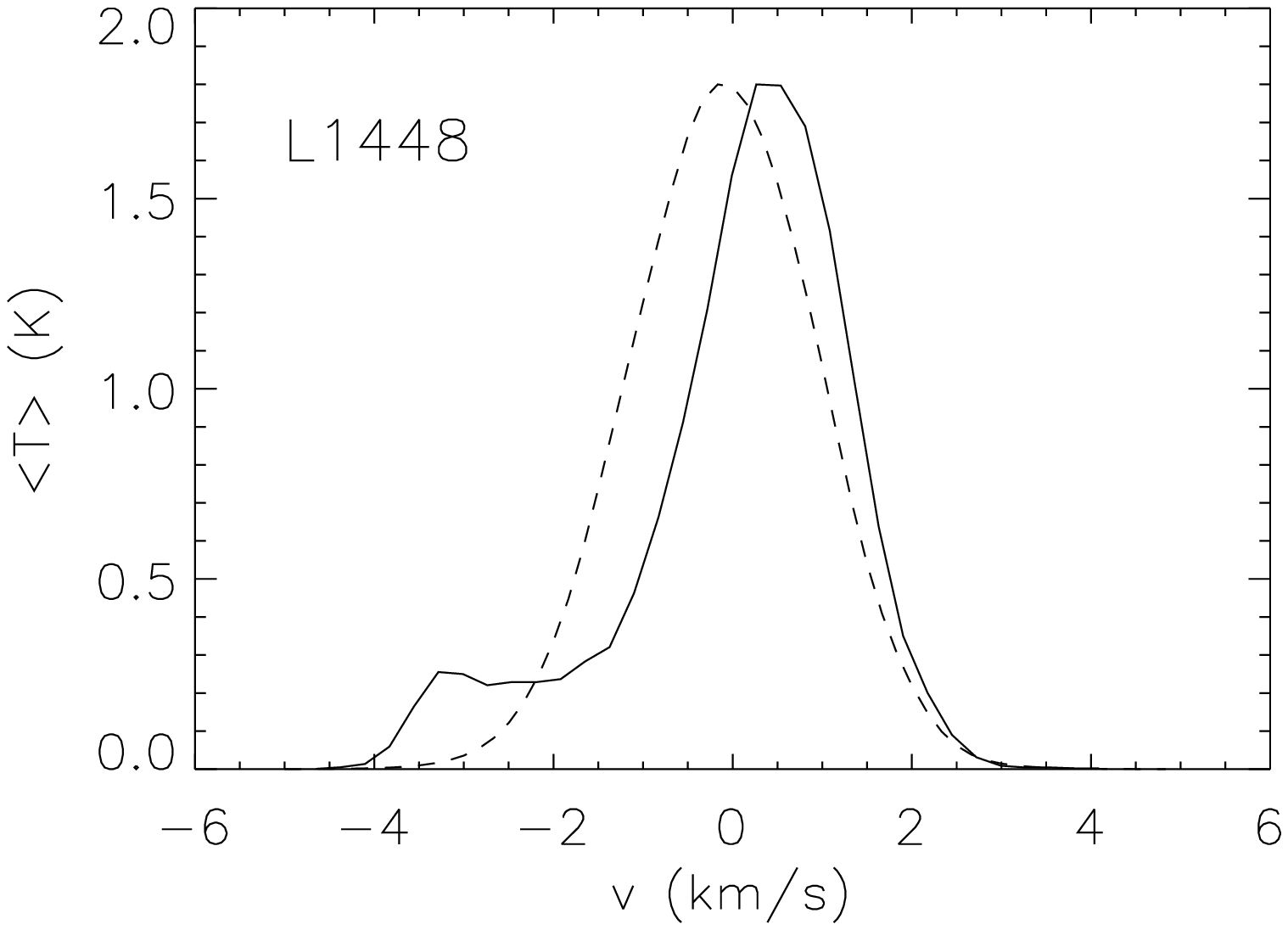}
\epsfxsize=0.5
\columnwidth
\epsfbox{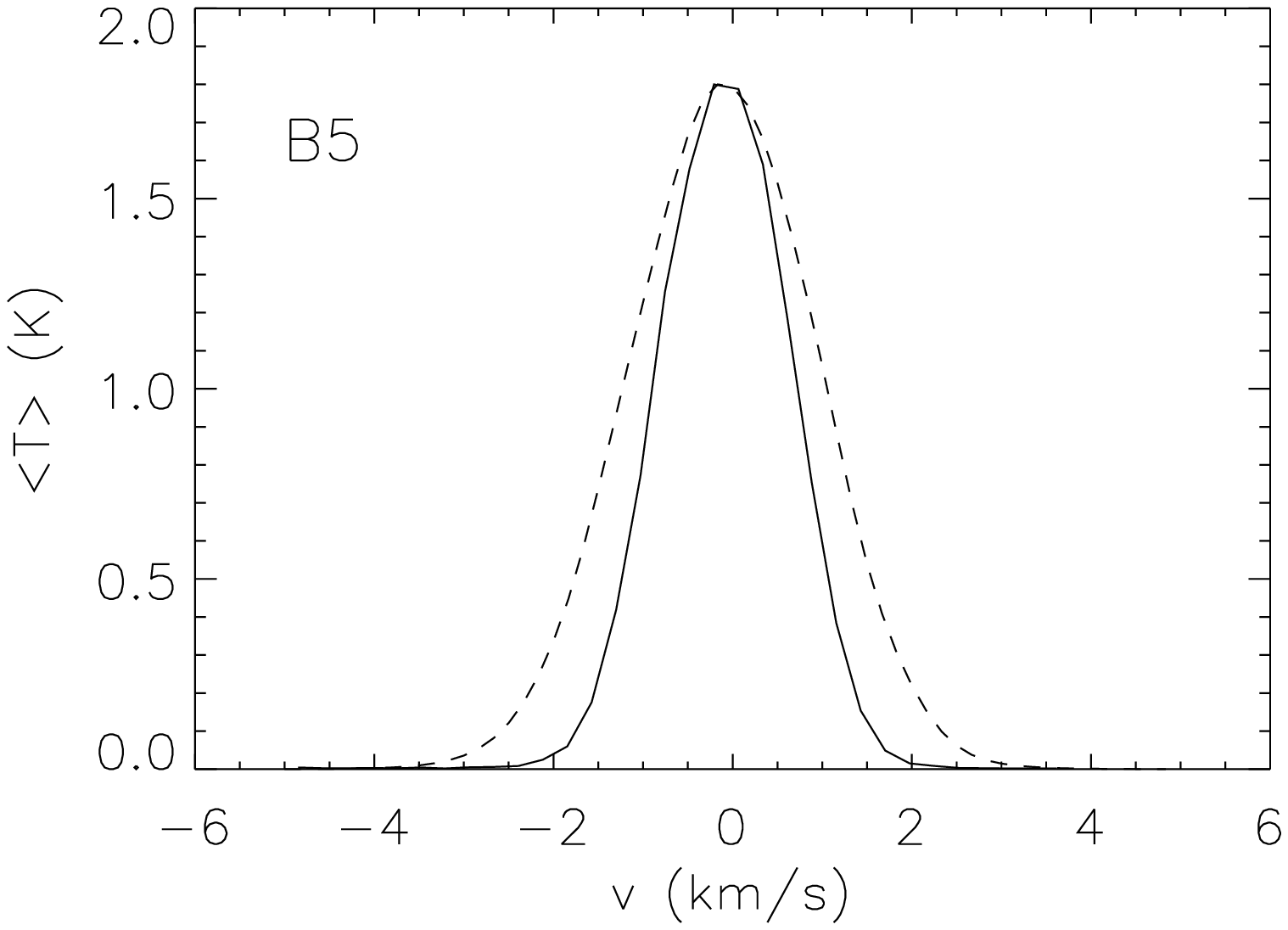}
\epsfxsize=0.5
\columnwidth
\epsfbox{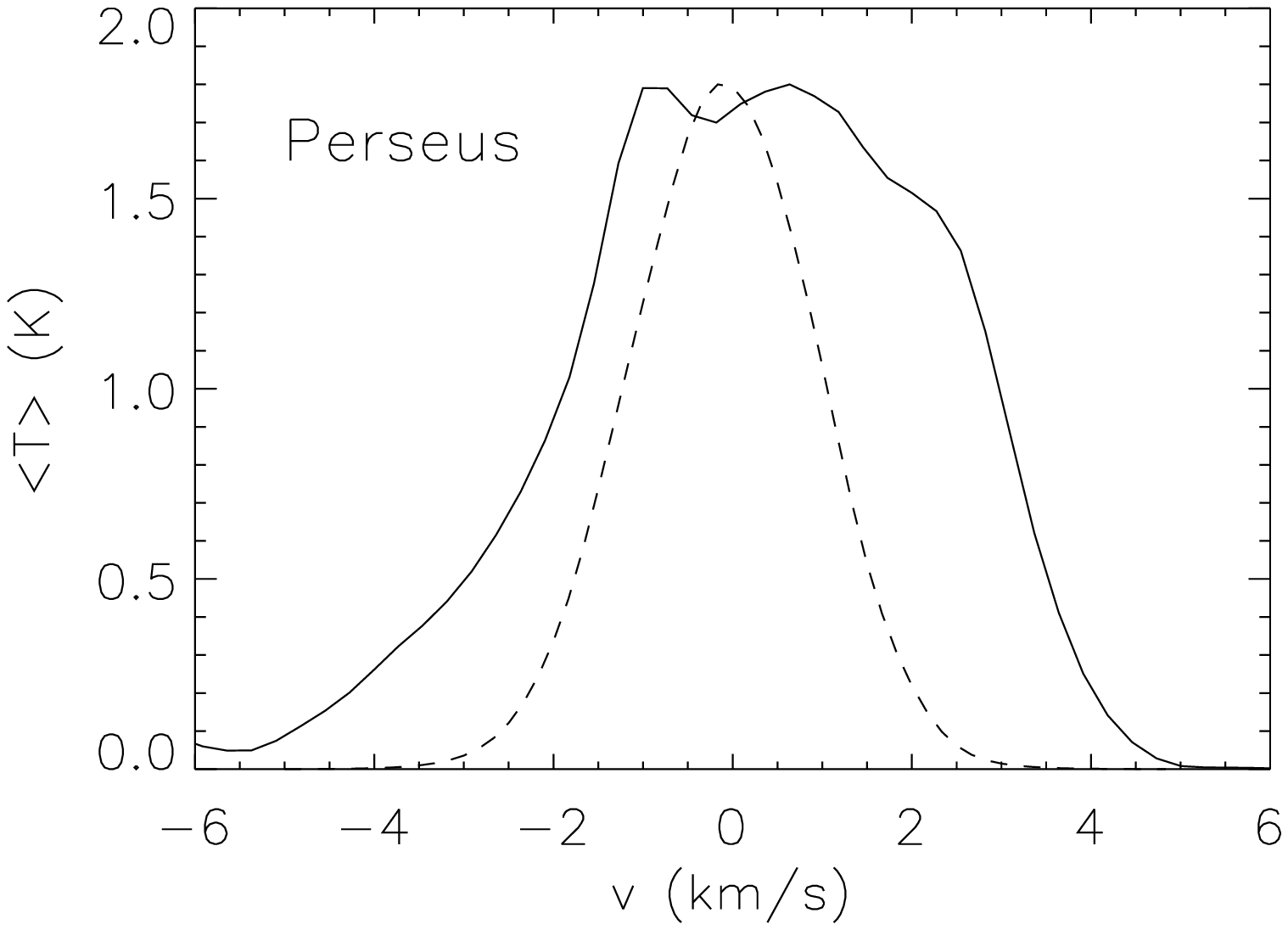}
\caption[]{}
\end{figure}

\newpage
\begin{figure}
\centering
\epsfxsize=0.5
\columnwidth
\epsfbox{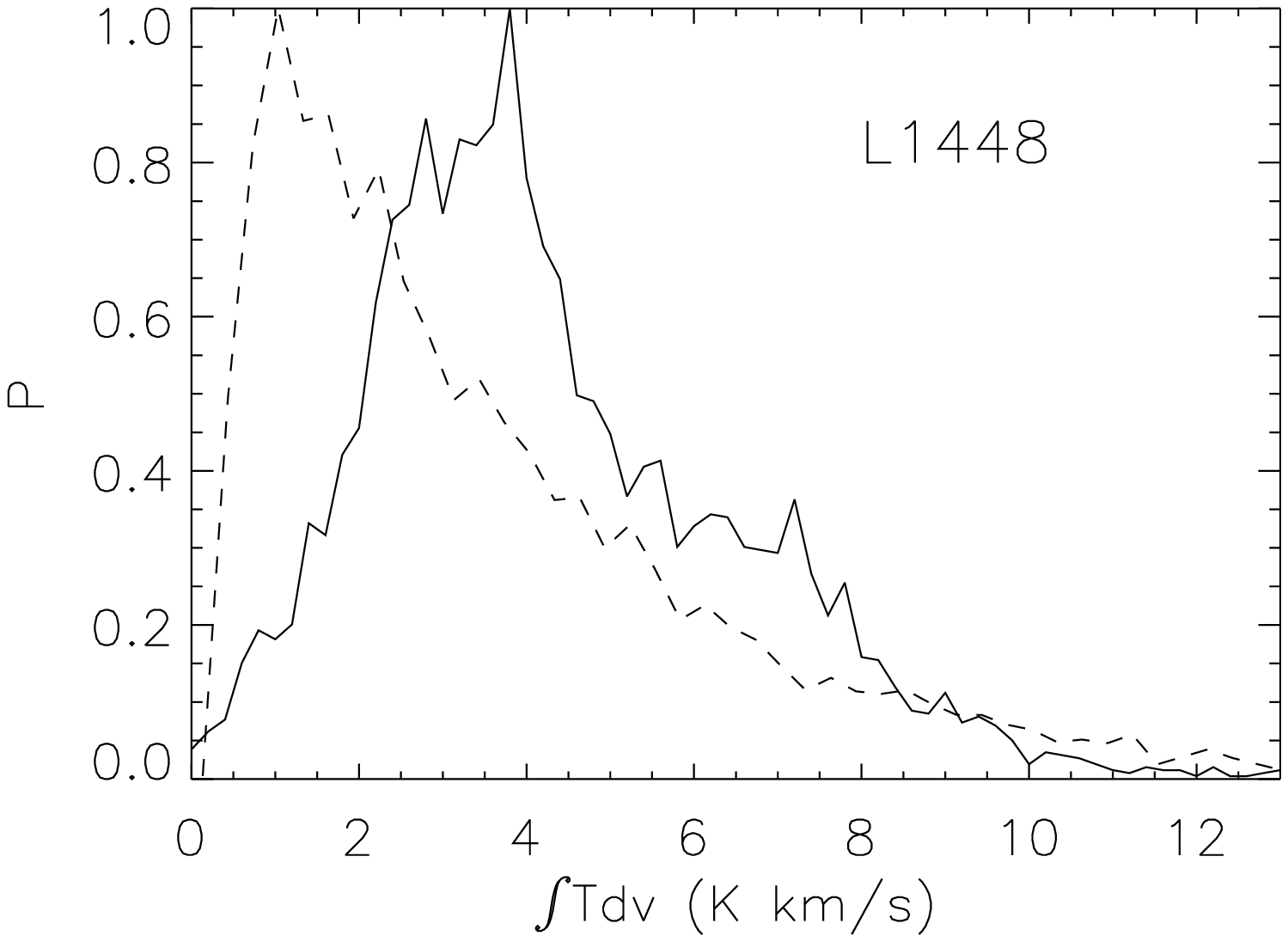}
\epsfxsize=0.5
\columnwidth
\epsfbox{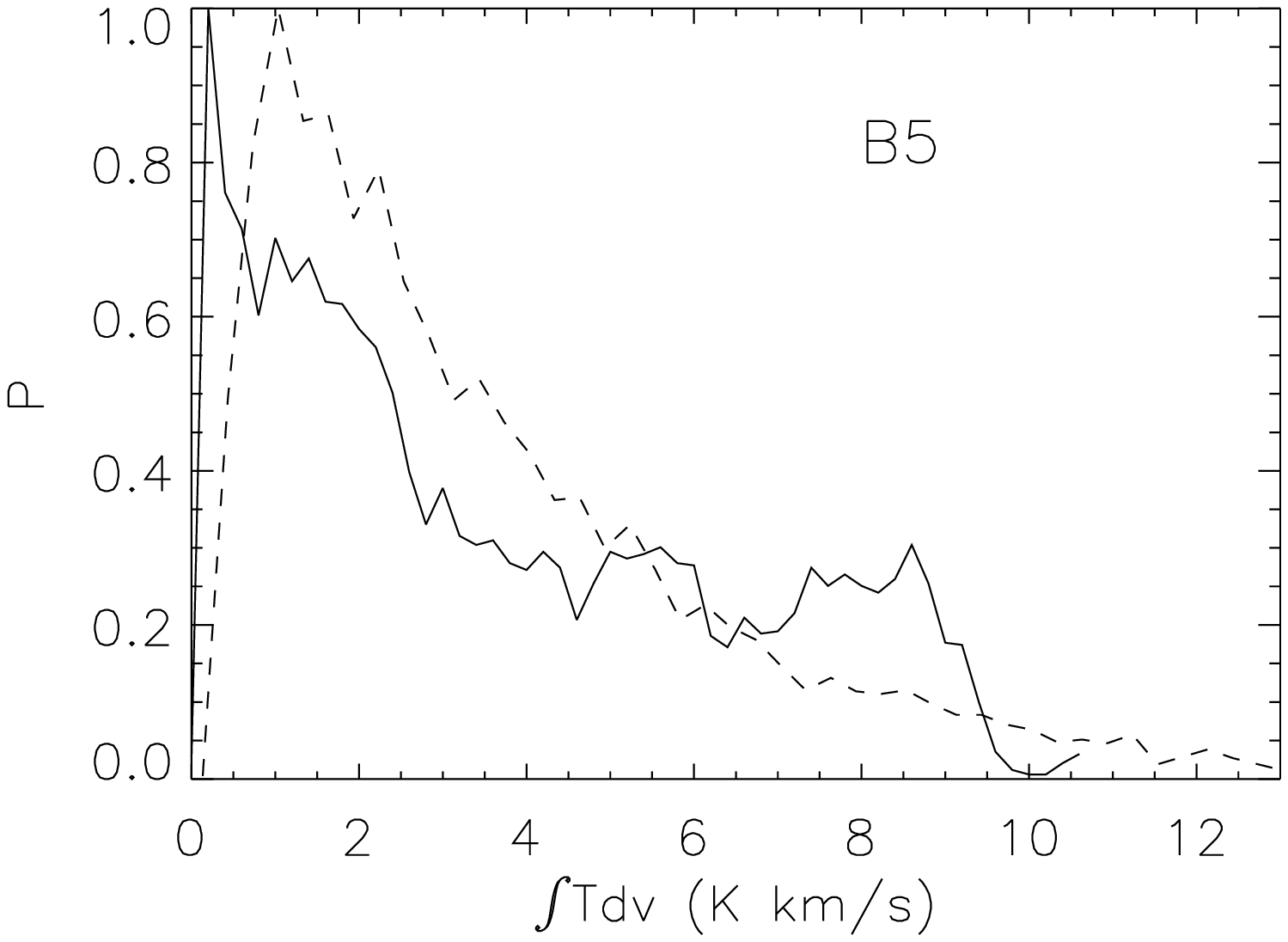}
\epsfxsize=0.5
\columnwidth
\epsfbox{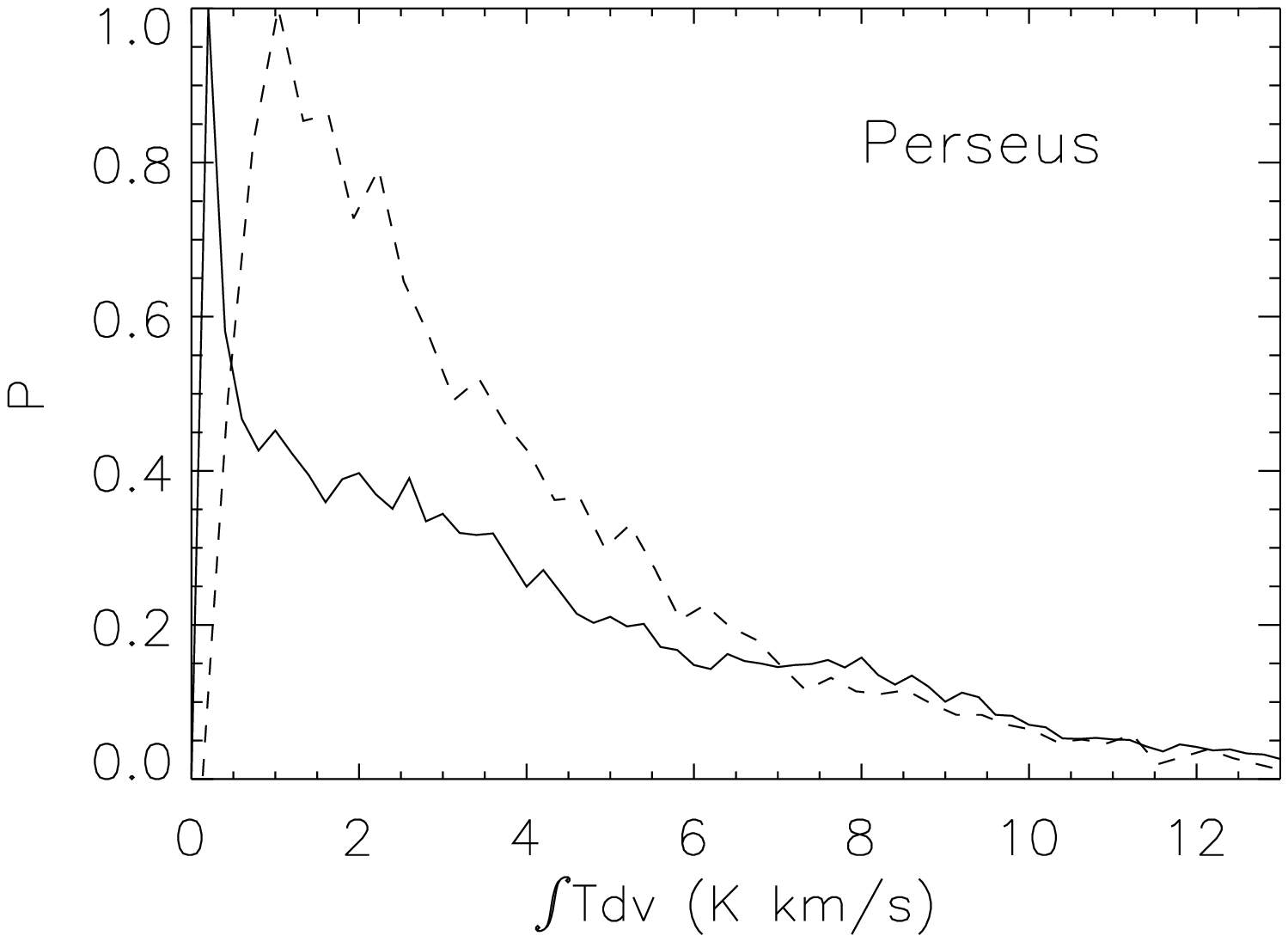}
\caption[]{}
\end{figure}

\newpage
\begin{figure}
\centering
\leavevmode
\epsfxsize=1.0
\columnwidth
\epsfbox{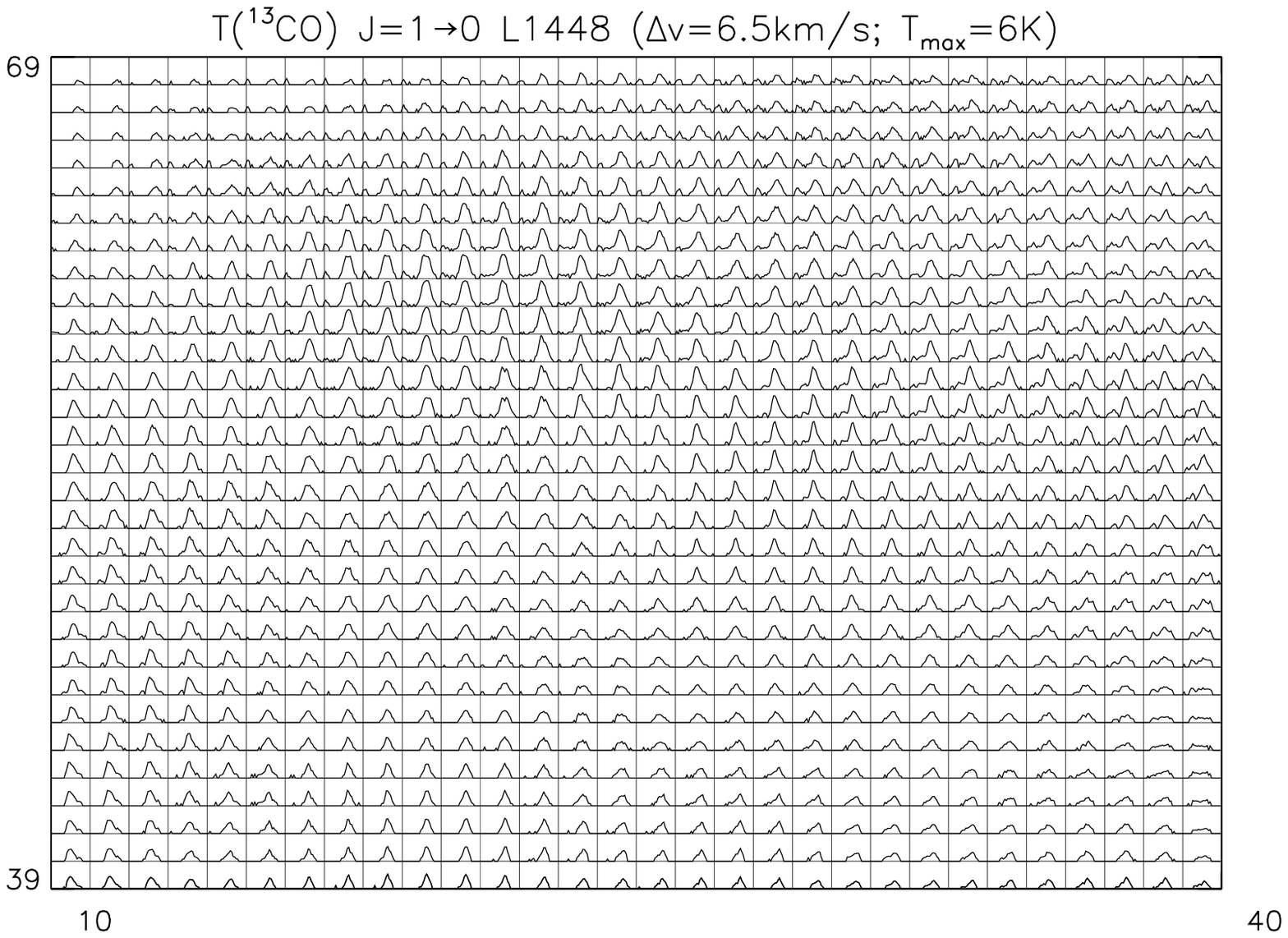}
\caption[]{}
\end{figure}

\newpage
\begin{figure}
\centering
\leavevmode
\epsfxsize=1.0
\columnwidth
\epsfbox{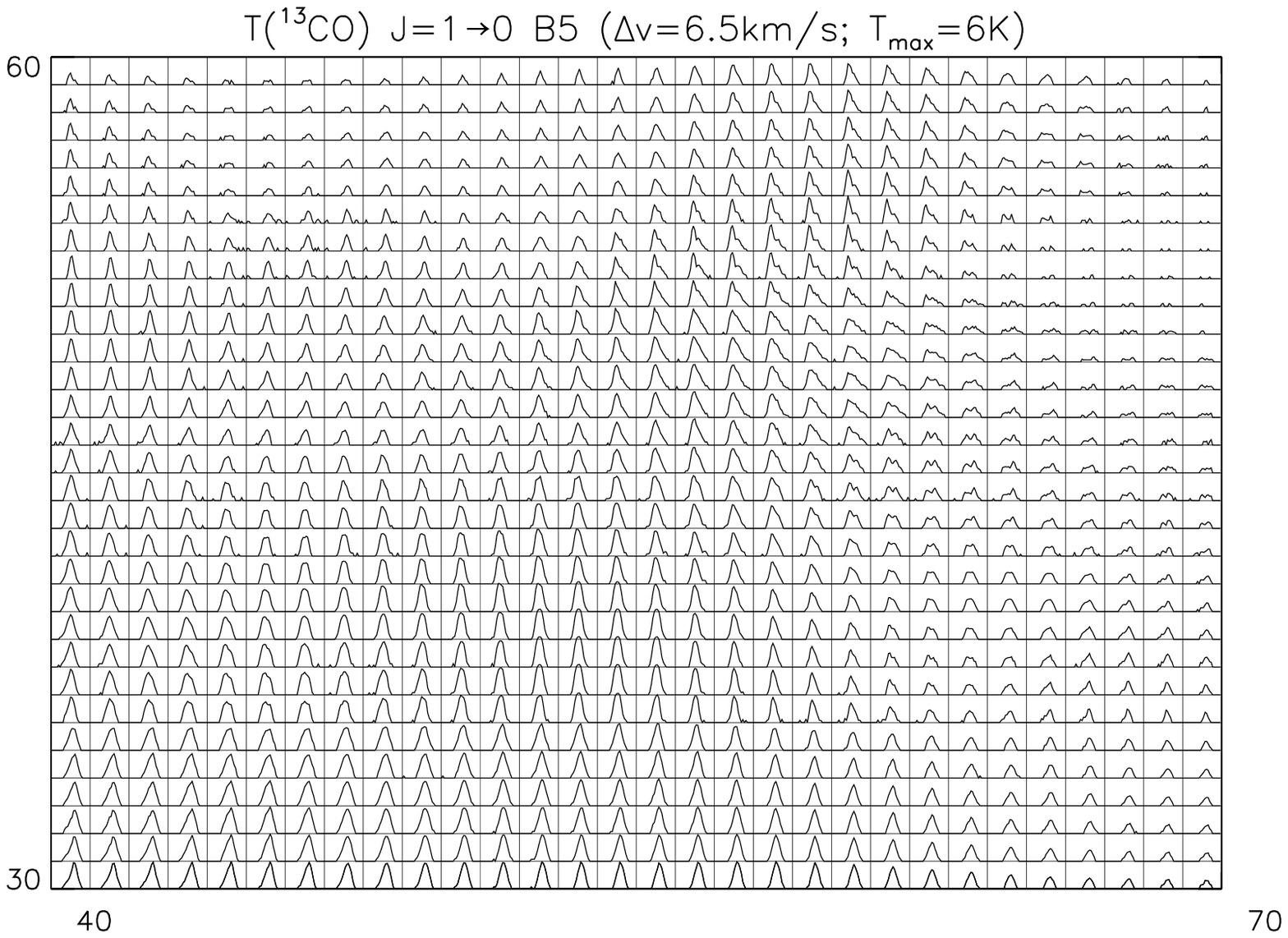}
\caption[]{}
\end{figure}

\newpage
\begin{figure}
\centering
\leavevmode
\epsfxsize=1.0
\columnwidth
\epsfbox{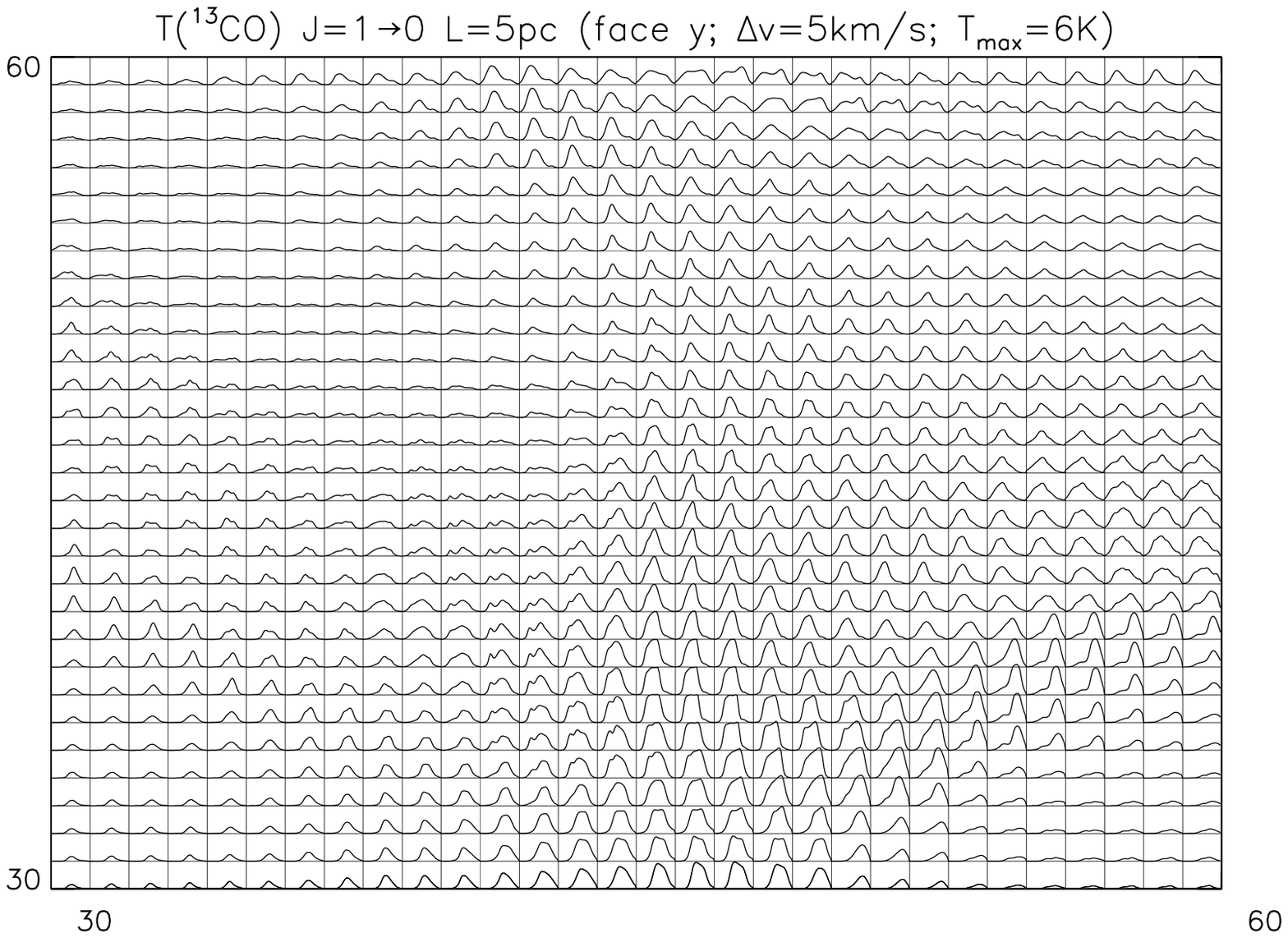}
\caption[]{}
\end{figure}

\newpage
\begin{figure}
\centering
\leavevmode
\epsfxsize=1.0
\columnwidth
\epsfbox{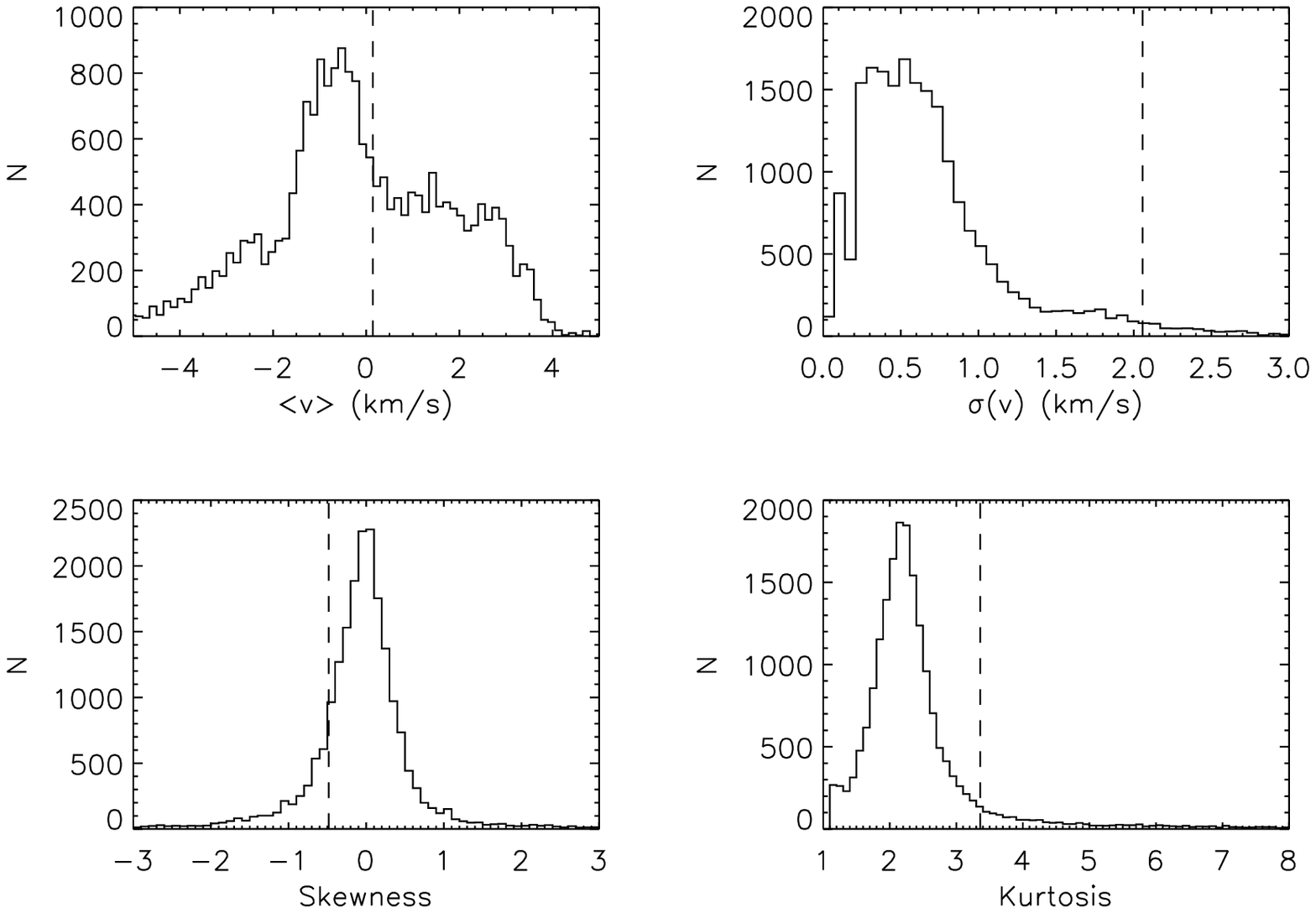}
\caption[]{}
\end{figure}

\newpage
\begin{figure}
\centering
\leavevmode
\epsfxsize=1.0
\columnwidth
\epsfbox{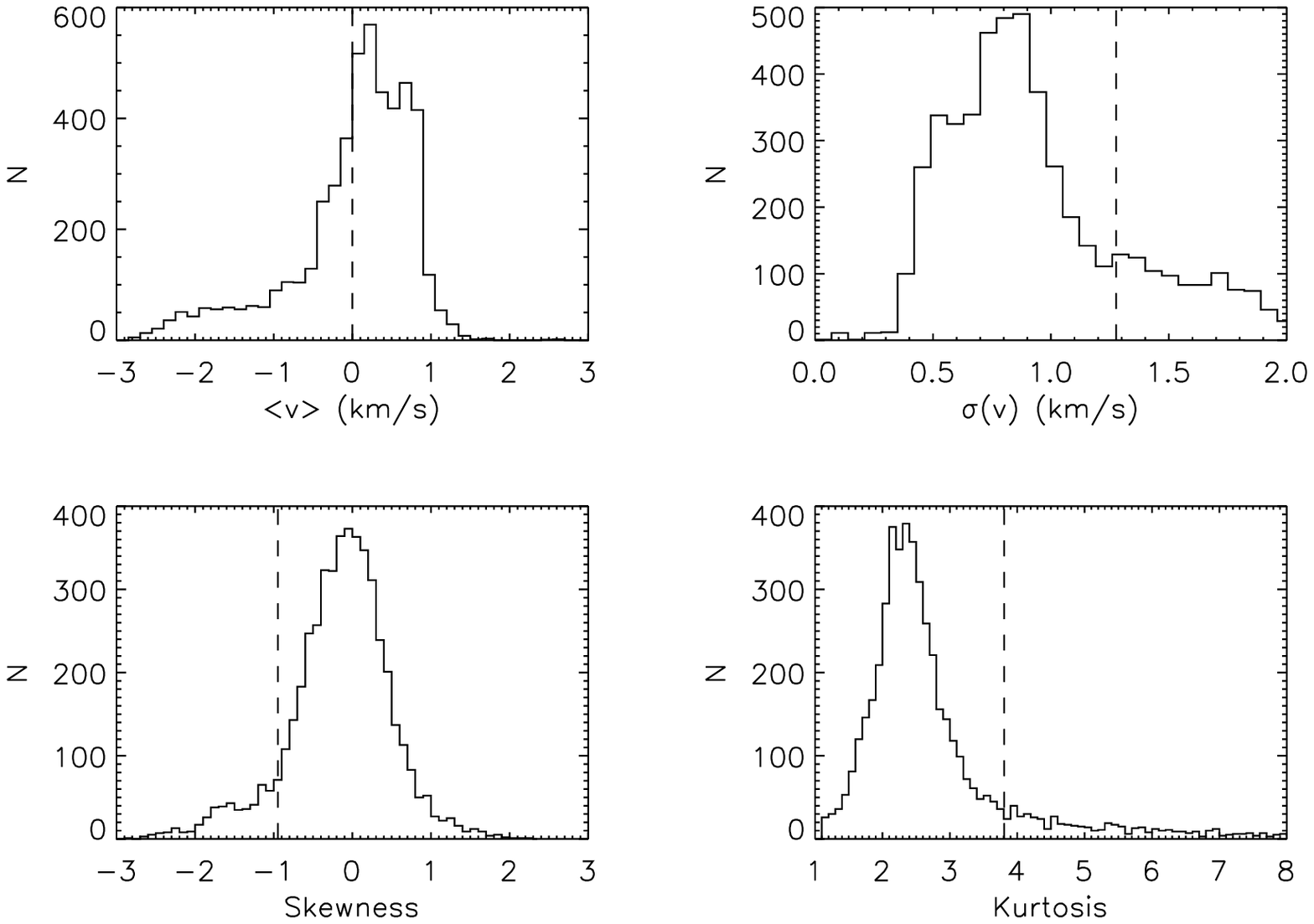}
\caption[]{}
\end{figure}

\newpage
\begin{figure}
\centering
\leavevmode
\epsfxsize=1.0
\columnwidth
\epsfbox{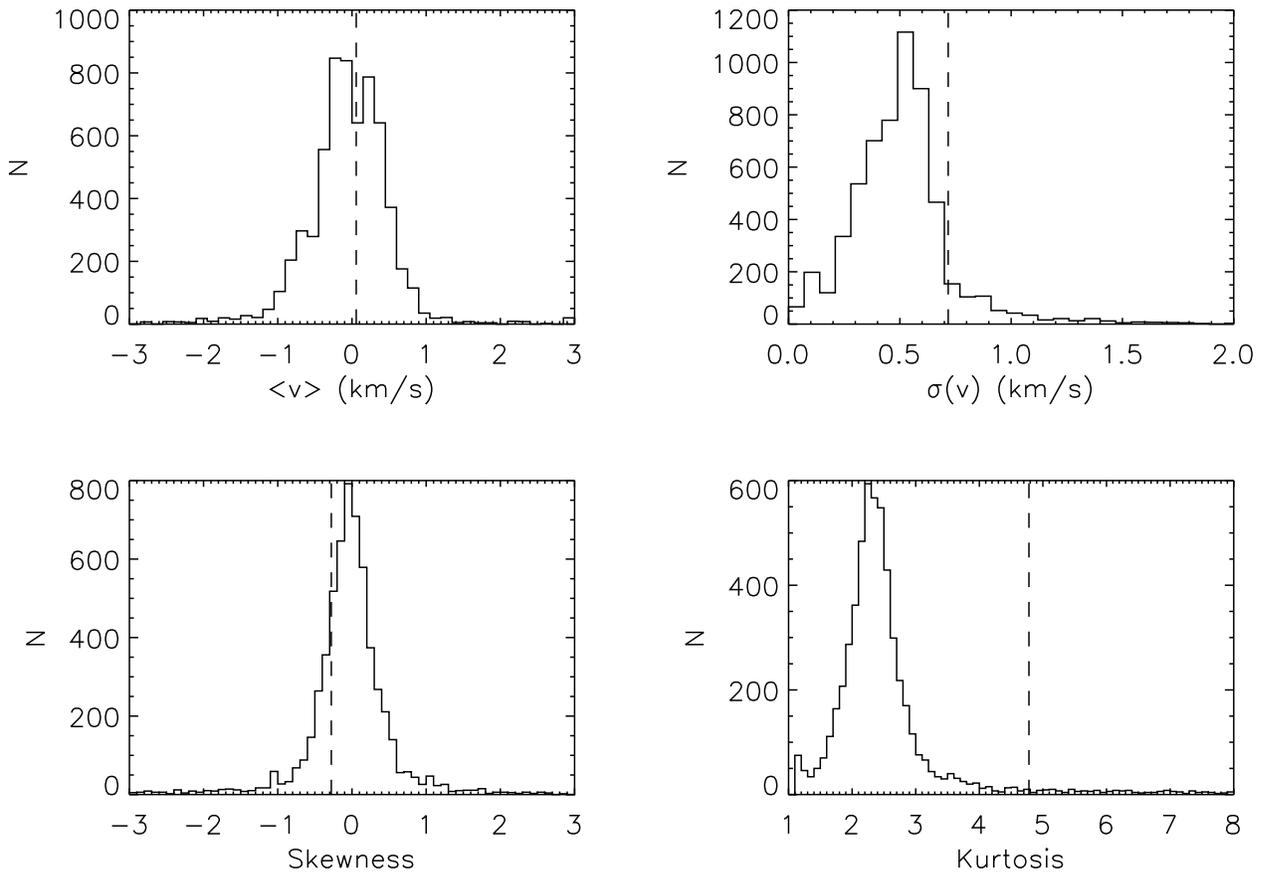}
\caption[]{As in figure 8 but for B5.}
\end{figure}

\newpage
\begin{figure}
\centering
\leavevmode
\epsfxsize=1.0
\columnwidth
\epsfbox{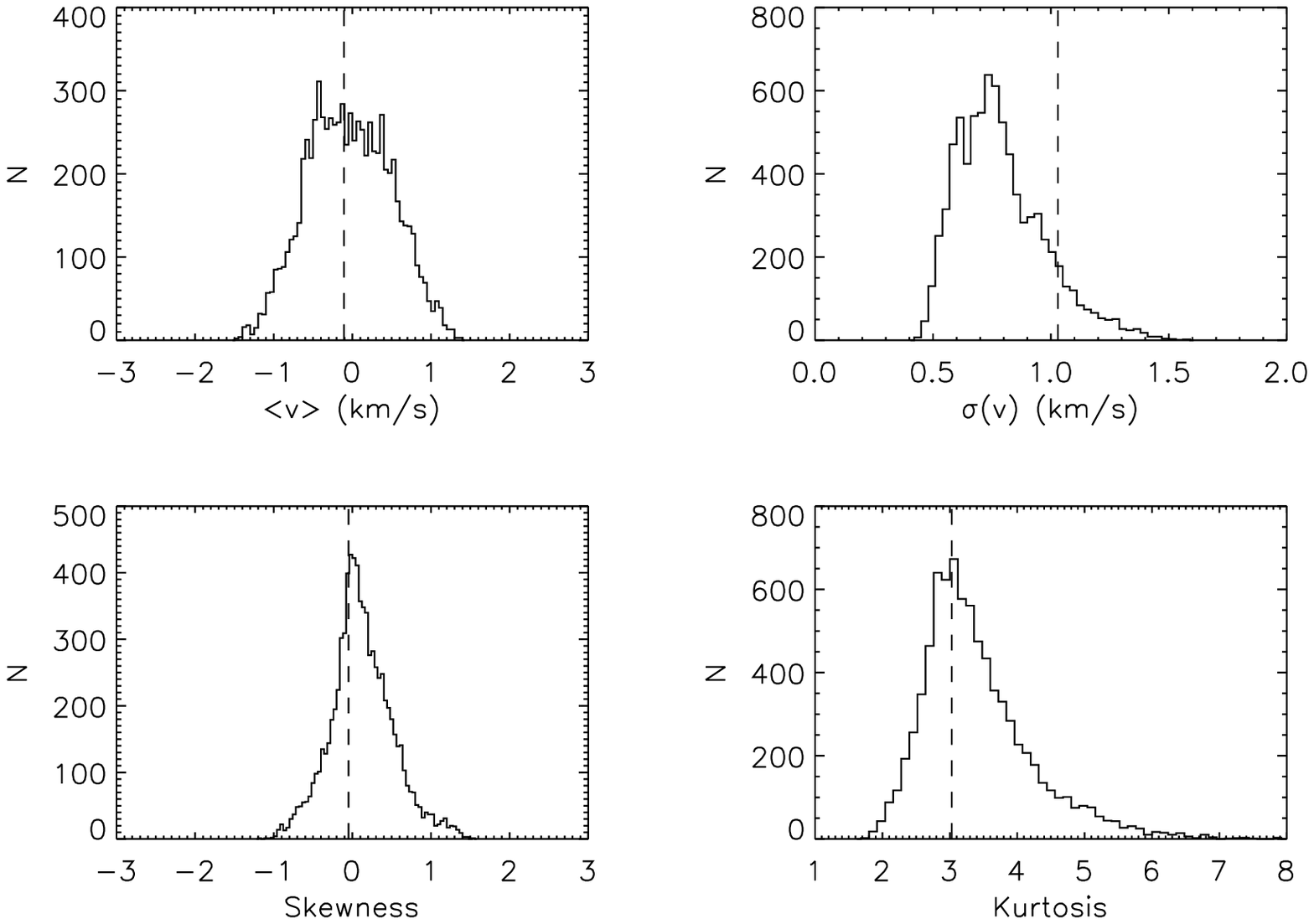}
\caption[]{}
\end{figure}

\clearpage
\newpage
\begin{figure}
\centering
\epsfxsize=0.45
\columnwidth
\epsfbox{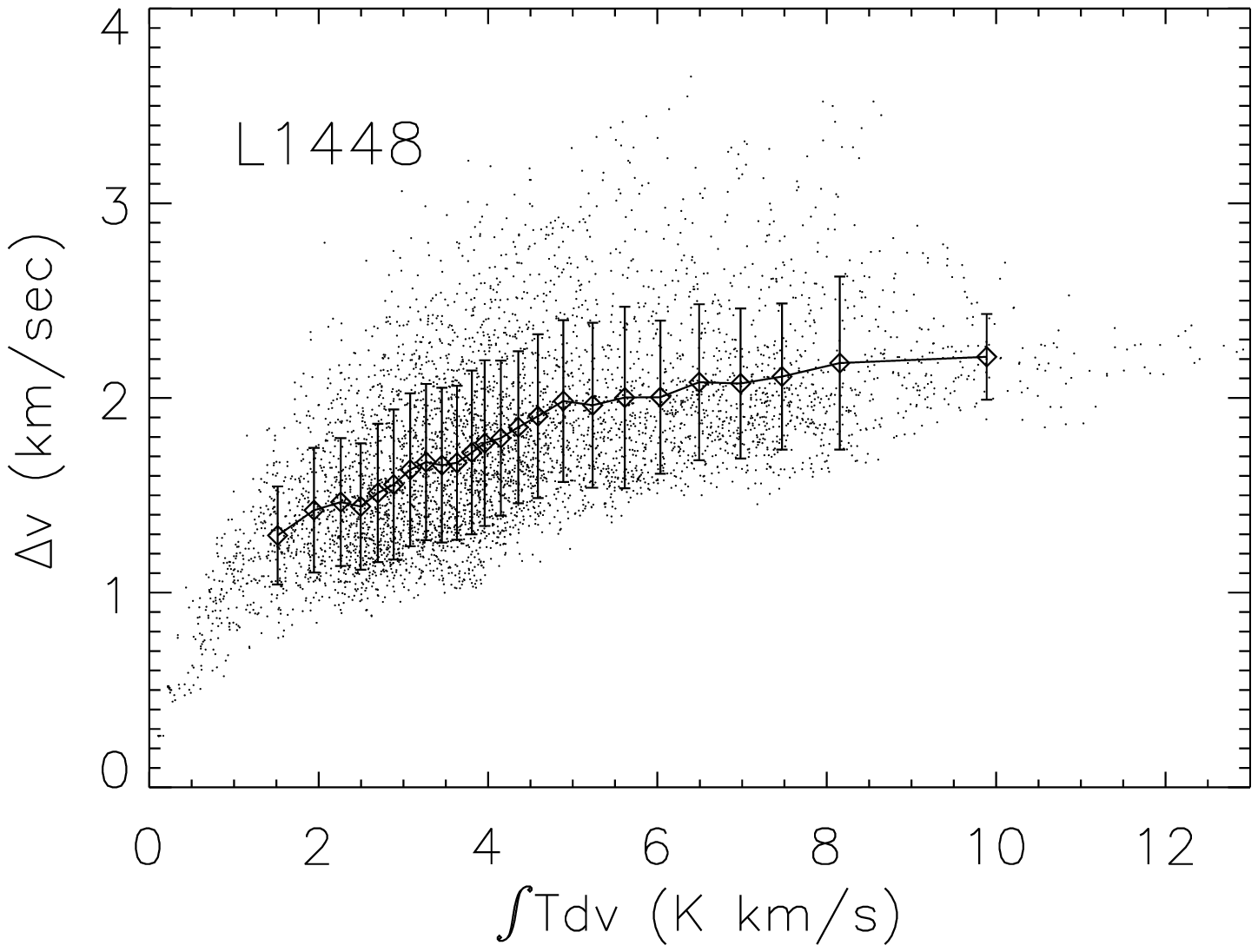}
\epsfxsize=0.45
\columnwidth
\epsfbox{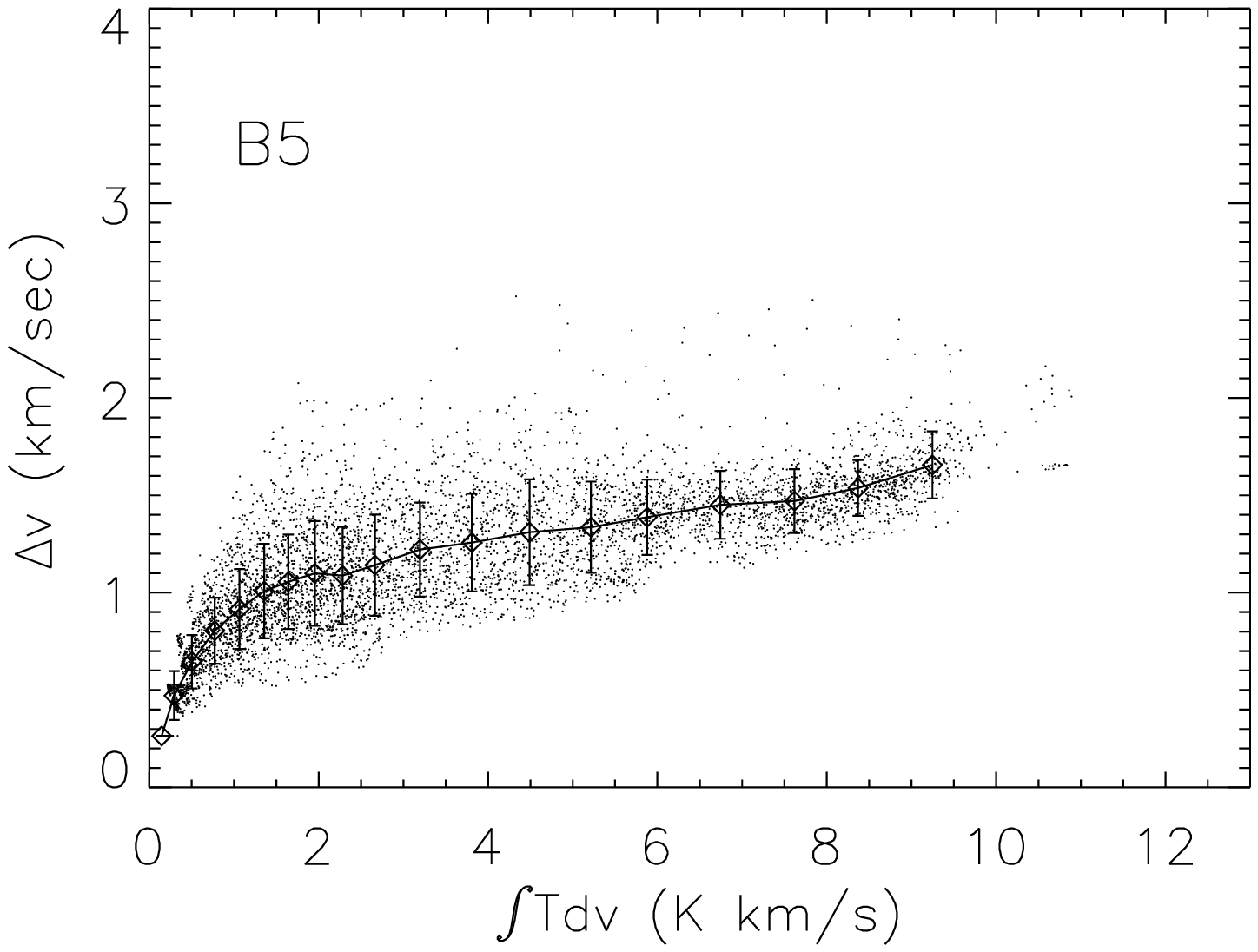}
\epsfxsize=0.45
\columnwidth
\epsfbox{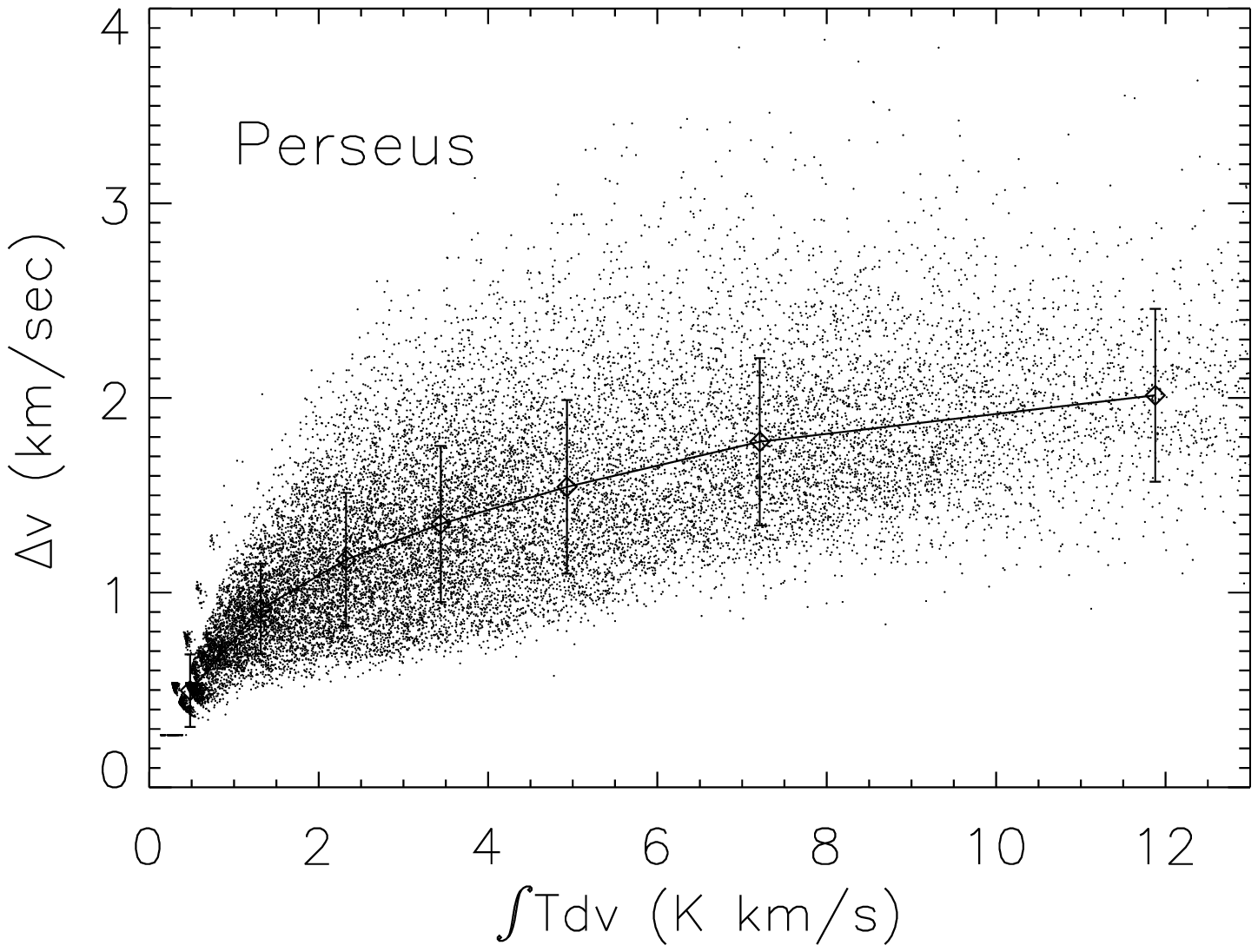}
\epsfxsize=0.45
\columnwidth
\epsfbox{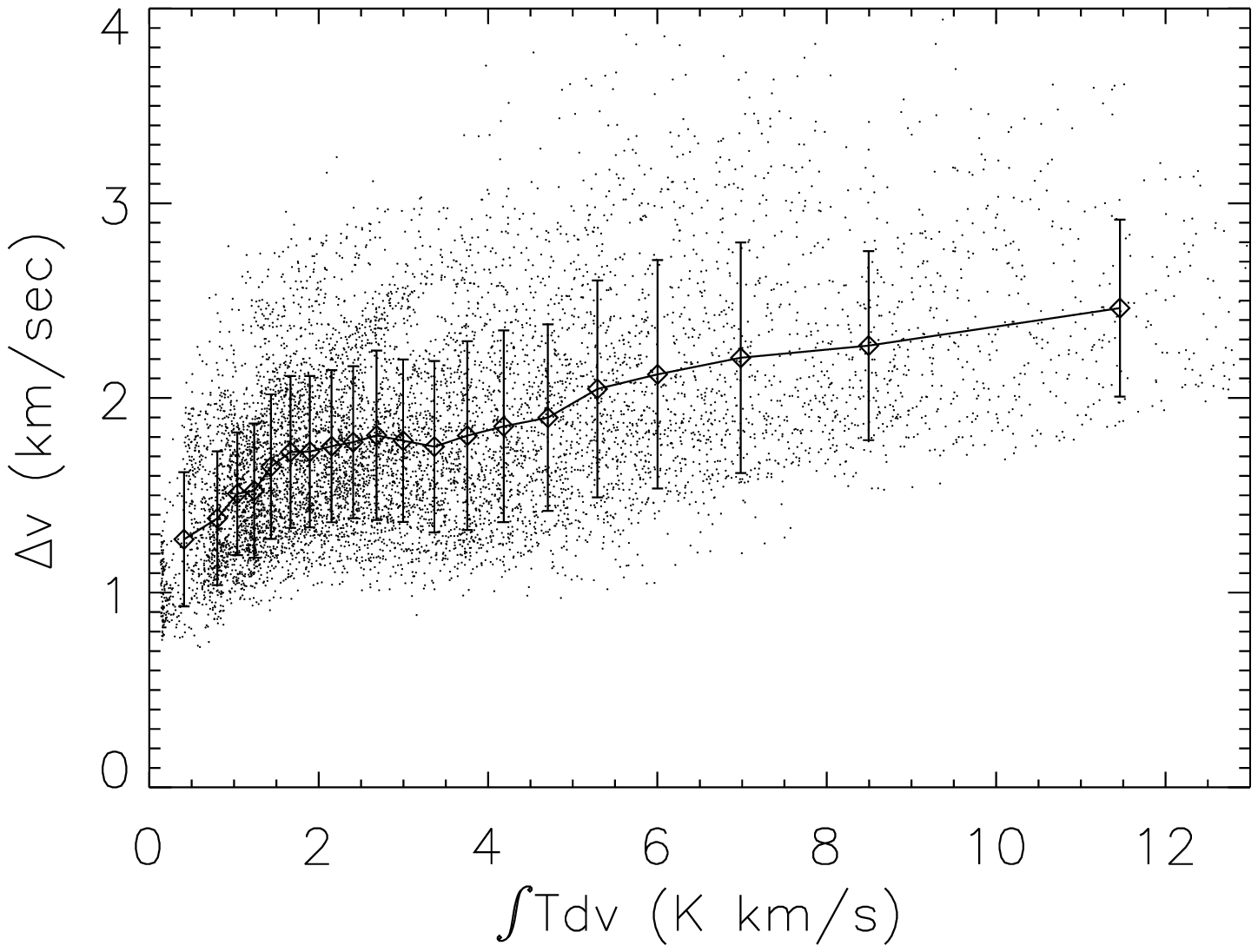}
\caption[]{}
\end{figure}

\end{document}